\documentstyle[amssymb,psfig]{mn2e}

\title[The initial mass distribution of the M82 star cluster
system]{The initial mass distribution of the M82 star cluster system}

\author[R. de Grijs, G. Parmentier and H.J.G.L.M. Lamers]{R. de
Grijs$^1$\thanks{E-mail: R.deGrijs@sheffield.ac.uk}, G. Parmentier$^2$
and H.J.G.L.M. Lamers$^{3,4}$ \\ 
$^1$ Department of Physics \& Astronomy, The University of Sheffield,
Hicks Building, Hounsfield Road, Sheffield S3 7RH\\ 
$^2$ Institute of Astronomy, University of Cambridge, Madingley Road,
Cambridge CB3 0HA\\ 
$^3$ Astronomical Institute, Utrecht University, Princetonplein 5,
3584 CC Utrecht, The Netherlands\\
$^4$ SRON Laboratory for Space Research, Sorbonnelaan 2, 3584 CC
Utrecht, The Netherlands}

\date{Received date; accepted date}
\pubyear{2005}

\begin{document}
\maketitle

\begin{abstract}
We explore whether we can constrain the shape of the {\it initial}
mass distribution of the star cluster population in M82's $\sim 1$
Gyr-old post-starburst region ``B'', in which the present-day cluster
mass function (CMF) is closely approximated by a log-normal
distribution. We conclude that the M82 B initial CMF must have had a
mean mass very close to that of the ``equilibrium'' CMF of Vesperini
(1998). Consequently, if the presently observed M82 B CMF has remained
approximately constant since its formation, as predicted, then the
{\it initial} CMF must have been characterized by a mean mass that was
only slightly larger than the present mean mass. From our detailed
analysis of the expected evolution of CMFs, we conclude that our
observations of the M82 B CMF are inconsistent with a scenario in
which the 1 Gyr-old cluster population originated from an initial
power-law mass distribution. Our conclusion is supported by arguments
related to the initial density in M82 B, which would have been
unphysically high if the present cluster population were the remains
of an initial power-law distribution.
\end{abstract}


\section{Introduction}
\label{intro.sec}

The derivation of galaxy formation and evolution scenarios using their
star cluster systems as tracers is limited to the study of {\it
integrated} cluster properties for galaxies beyond the Magellanic
Clouds, even at {\sl Hubble Space Telescope} spatial resolution. In
this context, one of the most important and most widely used
diagnostics is the distribution of cluster luminosities, or --
alternatively -- their associated masses, commonly referred to as the
cluster luminosity and mass functions (CLF, CMF), respectively.

In de Grijs et al. (2003b; see also de Grijs 2002; de Grijs et
al. 2003a) we reported the discovery of an approximately Gaussian (or
log-normal) CLF (and CMF) for the roughly coeval star clusters at the
intermediate age of $\sim 1$ Gyr in M82's fossil starburst region B.
This provided the first deep CLF (CMF) for a star cluster population
at intermediate age, which thus serves as an important benchmark for
theories of the evolution of star cluster systems. Recently,
Goudfrooij et al. (2004) added a second important data point to
constrain such theories, based on the roughly 3 Gyr-old cluster
population in NGC 1316, for which they also detected a clear turn-over
in their CLF\footnote{However, note that based on the published CLFs
and the discussion in Goudfrooij et al. (2004), the turn-over in the 3
Gyr-old metal-rich ($Z \sim$ Z$_\odot$) ``inner'' cluster population
($R \le 9.4$ kpc) in NGC 1316 occurs at $M_V \sim -6.2$, with a half
width at half maximum (based on their Fig. 3f) of $\sim 1.2$ mag (FWHM
$\sim 2.4$ mag). Assuming a Salpeter-like IMF with masses $m_\ast \ge
0.1$ M$_\odot$, the {\sc galev} simple stellar population models
(Schulz et al. 2002; Anders \& Fritze-v. Alvensleben 2003) indicate a
mean cluster mass of $\log(M_{\rm cl}/{\rm M}_\odot) \sim 4.0$, with a
FWHM of $\sim 0.9$ dex. These are significantly smaller masses (and a
smaller width) than expected for globular cluster progenitors.}.

Starting with the seminal work by Elson \& Fall (1985) on the young
Large Magellanic Cloud (LMC) cluster system (with ages $\lesssim 2
\times 10^9$ yr) seems to imply that the CLF of young star clusters
(YSCs) is well described by a power law of the form $N_{\rm YSC}(L)
{\rm d} L \propto L^{\alpha} {\rm d} L$, where $N_{\rm YSC}(L) {\rm d}
L$ is the number of YSCs with luminosities between $L$ and $L + {\rm
d} L$, with $-2 \lesssim \alpha \lesssim -1.5$ (e.g., Elmegreen \&
Efremov 1997; Whitmore et al. 2002; Bik et al. 2003; de Grijs et
al. 2003c; Hunter et al. 2003; see also Elmegreen 2002). On the other
hand, for old globular cluster (GC) systems with ages $\gtrsim
10^{10}$ yr, the CLF shape is well established to be roughly Gaussian
(or log-normal), characterized by a peak (turn-over) magnitude at
$M_V^0 \simeq -7.4$ mag and a Gaussian FWHM of $\sim 3$ mag (Harris
1991; Whitmore et al. 1995; Harris et al. 1998). This shape is almost
universal, showing only a weak dependence on the metallicity and mass
of the host galaxy, and on the position within the galaxy (e.g.,
Harris 1996; Gnedin 1997; Kavelaars \& Hanes 1997; Baumgardt 1998;
Whitmore et al. 2002; Dirsch, Schuberth \& Richtler 2005).

This type of observational evidence has led to the popular, but thus
far mostly speculative theoretical prediction that not only a
power-law, but {\it any} initial CLF (CMF) will be rapidly transformed
into a Gaussian (or log-normal) distribution because of (i) stellar
evolutionary fading of the lowest-luminosity (and therefore
lowest-mass) objects to below the detection limit; and (ii) disruption
of the low-mass clusters due both to interactions with the
gravitational field of the host galaxy, and to internal two-body
relaxation effects leading to enhanced cluster evaporation (e.g.,
Elmegreen \& Efremov 1997; Gnedin \& Ostriker 1997; Ostriker \& Gnedin
1997; Fall \& Zhang 2001).

The shape of the CLF (CMF) of YSC systems has recently attracted
renewed theoretical and observational attention. It has been pointed
out that for YSCs exhibiting an age range, one must first correct
their CLF to a common age before a realistic assessment of both their
present-day and {\it initial} CLF (CMF) can be achieved (e.g., Meurer
1995; Fritze-v. Alvensleben 1998, 1999; de Grijs et al. 2001,
2003a,b). Whether the observed power laws of the CLF and CMF for YSC
systems are intrinsic to the cluster population or artefacts caused by
the presence of an age spread in the cluster population -- which might
mask a differently shaped underlying distribution -- is therefore a
matter of ongoing debate (see, e.g., Fritze-v. Alvensleben 1998, 1999;
Carlson et al. 1998; Whitmore et al. 1999; Zhang \& Fall 1999;
Vesperini 2000, 2001).

The models built on either of these two assumptions for the initial
mass distributions, i.e., power law or log-normal, give rise to
temporal dependences of the CLF and CMF that are already well
established by the time a cluster population reaches the age of
$\lesssim 1$ Gyr (see, e.g., Vesperini 1998, 2000, 2001; Fall \& Zhang
2001), i.e., the age of the M82 B cluster population. The M82 B
cluster population represents an ideal sample to test these
evolutionary scenarios for, since it is a roughly coeval
intermediate-age population in a spatially confined region, where the
characteristic cluster disruption time-scale is among the shortest
known in any galactic disc region (e.g., de Grijs et al. 2003a, but
see Section \ref{m82tdis.sec}).

In this paper (Section \ref{models.sect}) we will compare the
observationally determined parameters for the M82 CLF and CMF
(introduced and discussed in Section \ref{m82to.sect}) to the model
predictions based on both the log-normal (Vesperini 1998, 2000, 2001)
and the initial power-law distributions (e.g., Fall \& Zhang 2001). We
will then (Section \ref{m82tdis.sec}) discuss the best constraints we
can set on the disruption time-scale for M82, assuming both an initial
log-normal distribution and an initial power law (following de Grijs
et al. 2003a). This will then be used (Section \ref{discussion.sec})
to arrive at our best estimate for the shape of the initial CMF in M82
B. Finally, we summarise our results and conclusions in Section
\ref{conclusions.sect}.

\section{The M82 B cluster mass function}
\label{m82to.sect}

We detected some 110 young and intermediate-age star clusters in the
post-starburst region ``B'' near the centre of the nearby, prototype
starburst galaxy M82 (de Grijs et al. 2001). Their age distribution
showed a clear peak around the time of the onset of the gravitational
interaction with M82's large neighbour spiral, M81.

Our recent re-analyses of these data using an improved approach (de
Grijs et al. 2003a,b, Parmentier, de Grijs \& Gilmore 2003) confirmed
that the M82 B cluster system is characterized by a significant,
well-defined peak of cluster formation, roughly defined within the age
limits of $8.7 \lesssim \log ( {\rm Age/yr} ) \lesssim 9.2$. Since
this peak may have been broadened by uncertainties in the age
determinations (see de Grijs et al. 2003a), this age range should be
considered an upper limit to the duration of enhanced cluster
formation in M82 B. This implies that the clusters contained in this
peak likely represent a roughly coeval population. For such a coeval
population, the observational selection effects are very well
understood (de Grijs et al. 2003a,b), while the dynamical cluster
disruption effects are very similar for this entire population.

\subsection{Robust detection of a log-normal CMF at intermediate age}

\begin{figure}
\psfig{figure=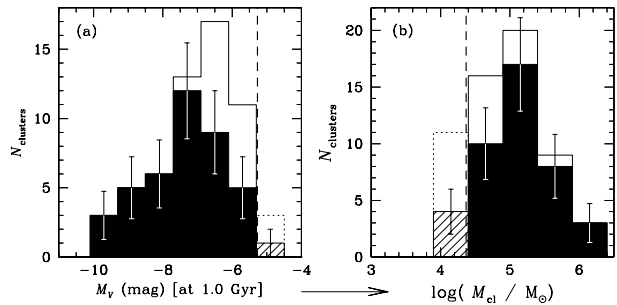,width=8.8cm}
\vspace{-4cm} 
\caption{\label{turnover.fig}Age-normalised CLF and corresponding mass
distribution of the M82 B clusters formed in the burst of cluster
formation, $8.7 \le \log ({\rm Age/yr}) \le 9.2$. The shaded
histograms correspond to the clusters with well-determined ages (see
de Grijs et al. 2003a); the open histograms represent the entire
cluster sample covering the age range of the burst. The vertical
dashed line is our 100 per cent completeness limit.}
\end{figure}

We restricted our analysis to the CLF, and its associated CMF, of the
clusters formed in the peak of cluster formation to avoid unnecessary
and ill-understood complications. We corrected our ``peak cluster
sample'' to a common age of 1 Gyr (i.e., coinciding with the peak of
cluster formation), using the Starburst99 SSP models (Leitherer et
al. 1999), although the effects of this correction are small because
of the relatively narrow age range. The resulting CLF at a common age
of 1 Gyr, and the corresponding CMF, are shown in Fig.
\ref{turnover.fig}, where we distinguish between clusters with
well-determined ages [i.e., $\Delta \log( {\rm Age/yr} ) \le 1.0$;
shaded histograms] and the full cluster sample (open histograms)
covering the age range of the burst.

Within the -- mostly Poissonian -- uncertainties, both the
age-normalised CLF and the CMF can be adequately described by a
log-normal (or Gaussian) distribution. The CLF in Fig.
\ref{turnover.fig}a is characterised by a peak luminosity of $M_V^0 =
-7.3 \pm 0.1$ mag, and a Gaussian FWHM of $\sim 3.1$ mag. The CMF
exhibits a peak at $\log (M_{\rm cl} / {\rm M}_\odot) = 5.1 \pm 0.1$
and a Gaussian FWHM of $\sim 1.2$ dex ($\sigma_{\rm Gauss} \approx
0.5$ dex).

The fact that we considered an approximately coeval subset of the M82
B cluster population, combined with our use of the 100 per cent
completeness limit as our base line ensures the robustness of the CMF
peak detection. The original sample selection was essentially based on
a cross correlation of the source detections in the $V$ (F555W) and
$I$ (F814) {\sl Hubble Space Telescope} filters, visually complemented
by extended objects that were missed by the automated source
detection. None of the latter objects, however, were brighter than the
100 per cent completeness limit of the data in M82 B. While in de
Grijs et al. (2001) we quoted the completeness limits for point
sources, we found that the equivalent limits for ``realistic'' cluster
sizes of $R_{\rm eff} = 5$ pc vary by $\lesssim 0.5$ mag. This is
sufficiently small so as to not affect our results on the turn-over
(considering that the vast majority of clusters are only slightly more
extended than point sources -- see fig. 10 in de Grijs et al. 2001),
which occurs some 2 mag brighter than the 100 per cent completeness
limit. 

One have to be aware that variable extinction across the field of view
(as present in M82 B; de Grijs et al. 2001, 2003a; Parmentier et
al. 2003) could in principle cause an artificial turn-over even if the
intrinsic luminosity distribution were a power law. However, we note
that most clusters are affected by $A_V \ll 1$ mag, with only a very
small fraction having $A_V > 1$ mag. Nevertheless, in order to account
for the variable extinction and other potential selection effects, we
did not use the 50 per cent completness limit to base our conclusions
on regarding the turn-over (as is customarily done), but the full, 100
per cent limit. The difference between these 2 limits is $\sim1$ mag
(see fig. 7a in de Grijs et al. 2001), while the turn-over was found
another $\sim2$ mag brighter than this limit. Unless the extinction
for most clusters is well above $A_V = 1$ mag, (which is not supported
by our analysis), this essentially rules out extinction effects as
cause for an artificial turn-over.

On a related note, since all of the clusters considered for the
turn-over analysis have approximately the same age, it follows from
stellar population synthesis that they are also characterised by
similar colours, so that {\it any} extinction will have a blanket
effect on all clusters; variable extinction will dim different
clusters differently, but will not in any way be colour related for
this particular coeval sample.

Finally, the most important assumption we made to validate the
turn-over as based on a statistically complete sample is that the star
clusters we detected, i.e., the ones relatively close to the surface
area of M82 B, are fully representative of the M82 B population as a
whole. We believe this to be justified, for the following reason. The
population of M82 B clusters peaks at an age of $\sim 1$ Gyr. Over
such a period of time since their formation, differential rotation of
the region in which they were found, between about 0.5 and 1 kpc from
the galaxy's centre, would have been expected to have smeared out the
clusters' locations, yet this has not happened. This is most likely
owing to the fact that M82 B is spatially coincident with the end of
the M82 bar (see de Grijs 2001 for a discussion). M82 B has therefore
retained its intrinsic properties over at least the last Gyr -- this
applies to both the surface and the interior of the region. As such,
we believe that the clusters we discuss here are representative of the
area as a whole.

\subsection{Luminosity to mass conversion: choice of IMF}

Our M82 B cluster mass estimates are based on a Salpeter-like stellar
IMF (with masses $m_\ast \ge 0.1$ M$_\odot$). However, we realise that
recent determinations of the stellar IMF deviate significantly from
that representation at low masses. The low-mass stellar IMF is
significantly flatter than the Salpeter slope. The implication of this
is, therefore, that we have {\it overestimated} the individual cluster
masses. If we use a more modern IMF, such as that of Kroupa, Tout \&
Gilmore (1993), we have likely overestimated our individual cluster
masses by a factor of 1.7 to 3.5 for an IMF containing stellar masses
in the range $0.1 \le m_\ast/{\rm M}_\odot \le 100$ (de Grijs et
al. 2003a). The exact factor depends on which slope we adopt for the
lowest stellar mass range, $0.08 < m_\ast/{\rm M}_\odot \le 0.5$. This
corresponds to a correction of $-0.23$ to $-0.54$ in the peak of the
CMF, so that a more realistic estimate for the peak of the M82 B CMF
would be $\langle \log(M_{\rm cl}/{\rm M}_\odot) \rangle = 4.7 \pm
0.2$. The width of the corrected CMF remains the same, since every
individual cluster mass is simply corrected for by the same additive
amount in logarithmic mass space.

\section{Constraining the evolutionary scenarios}
\label{models.sect}

In de Grijs et al. (2003a,b) we also found that the characteristic
disruption time-scale for the clusters in M82 B, $t_{\rm dis}^4 \sim
30 \times 10^6$ yr for $10^4$ M$_\odot$ clusters, is considerably
shorter than that in the solar neighbourhood (i.e., $t_{\rm dis}^4
\sim 1.6 \times 10^9$ yr; Lamers et al. 2005b)\footnote{Note that this
short disruption time-scale is based on the assumption of an initial
power-law CLF (CMF). We will return to this issue in Section
\ref{m82tdis.sec}.}.

In the following subsections, we will discuss the implications of the
observed CLF and CMF shapes (i.e., the peaks and widths of the
distributions) at the associated intermediate age in the evolutionary
framework of both the initial power-law and the initial log-normal
mass distributions.

\subsection{Predictions for a power-law initial cluster mass 
distribution}

The most popular star cluster evolution models assume that the initial
distribution of cluster masses is well-represented by a power law,
which then rapidly transforms into a log-normal distribution due to
dynamical evolution effects (e.g., Harris \& Pudritz 1994; Okazaki \&
Tosa 1995; McLaughlin \& Pudritz 1996; Elmegreen \& Efremov 1997;
Gnedin \& Ostriker 1997; Murali \& Weinberg 1997; Vesperini 1997,
1998; Baumgardt 1998; Fall \& Zhang 2001, and references therein;
Parmentier \& Gilmore 2005; see also Elmegreen 2002). We note that
while there is general concensus that an initial power-law
distribution is rapidly transformed into a log-normal CMF, the
detailed predictions of these models differ significantly from each
other.

Whitmore et al. (2002) calculated the expected evolution of the
$\sim1.5$ Gyr-old log-normal-like CLF in NGC 3610, and showed that an
initial Schechter-type CMF [based on initial conditions and subsequent
evolution of the CMF given by Fall \& Zhang (2001)], in combination
with evolutionary fading of the stellar population, lead to the peak
luminosity and width of the cluster {\it luminosity} function to
remain virtually unchanged for a Hubble time. Model calculations based
on a power-law or Schechter-type initial cluster mass distribution
(e.g., Fall \& Zhang 2001) suggest that, in a Milky Way-type
gravitational potential with a strongly radially dependent radial
anisotropy of the cluster velocity distribution, the turn-over of the
cluster mass distribution will move towards higher masses by
approximately $\Delta \log (M_{\rm cl} / {\rm M}_\odot) \simeq +0.9$
by the time the cluster population reaches an age of 12 Gyr, i.e.,
similar to the median age of the Galactic GC system. This implies that
the star cluster system in M82 B, where the characteristic disruption
time-scale is significantly shorter than in the Galactic halo, will be
dominated by higher masses than the Galactic GC system when it reaches
a similar age, and most of the present-day clusters will be depleted.

Furthermore, Vesperini (2001) calculated the evolution of fiducial
star cluster systems with physically realistic parameters spanning the
entire observational parameter space, using {\it N}-body models, and
concluded that it is not straightforward, indeed very difficult, to
produce (nearly) universal CLFs and CMFs in very different types of
galaxies if starting from initial power-law distributions.

A first-order comparison between the observed CMF in M82 B (see
Fig. \ref{turnover.fig}b) and the predictions of the Fall \& Zhang
(2001) models also attests to these problems: while we observe a peak
in the M82 B CMF at $\log (M_{\rm cl} / {\rm M}_\odot) = 5.1 \pm 0.1$
and a Gaussian width consistent with the universal CMFs of old GC
systems, the Fall \& Zhang (2001) models at an age of 1.5 Gyr predict,
for either a power-law, a truncated power law (see also Parmentier \&
Gilmore 2005), or a Schechter-type initial CMF, a peak at $\log
(M_{\rm cl} / {\rm M}_\odot) \simeq 4.5$ and a significantly broader
mass distribution than observed for the Milky Way GC system (their
Fig. 3). Because of the systematic uncertainties inherent (i) to the
different methods of mass determination, (ii) to the sample
selection\footnote{We have shown (de Grijs et al. 2003a,b) that by
selecting a magnitude-limited cluster sample, our cluster sample is
almost 100 per cent complete (see Fig. 7 in de Grijs et al. 2001), so
that our results are very robust against statistical uncertainties.},
(iii) to the photometric uncertainties for our M82 B cluster sample,
and (iv) to the differences in the radial extent sampled [Fall \&
Zhang (2001) sample a large radial extent, at large galactocentric
distances, where the effects of dynamical friction are significantly
smaller than at the relatively small radii of our M82 B clusters],
these differences between the M82 B CMF and the model-predicted CMF of
Fall \& Zhang (2001) at a similar age may not be significant. We will
investigate the effects of different characteristic disruption
time-scales on the resulting CMF in Section \ref{discussion.sec}.

\subsection{A log-normal initial mass distribution?}

Vesperini (1998, 2000, 2001) suggested that the temporal evolution of
a {\it log-normal} initial CMF describes the currently observed old GC
luminosity and mass functions very well. Recently, Parmentier \&
Gilmore (2005) showed that in order to obtain the radial mass and
number distributions of the Milky Way's Old Halo GCs, the GC system
must have been depleted in low-mass objects {\it ab initio}. They
argue that the current CMF of Milky Way Old Halo GCs is most easily
obtained from an initial log-normal CMF, although a truncated
power-law CMF (truncated at $M_{\rm cl} \sim 10^5 {\rm M}_\odot$)
cannot be ruled out.

The rationale for adopting a log-normal initial CMF was provided by
the shape of the CMFs in the outer regions of massive elliptical
galaxies, where the initial conditions are likely retained because of
the low efficiency of cluster disruption processes expected at large
galactocentric distances (Vesperini 2000; see also McLaughlin, Harris
\& Hanes 1994; Gnedin 1997).

Vesperini's (1998, 2000) {\it N}-body models follow the evolution of
GC systems for a Hubble time in time-independent gravitational
potentials of Milky Way-type and ``elliptical'' host galaxies modeled
as isothermal spheres with constant circular velocity. His models
include the full treatment of stellar evolution (and hence mass loss
from individual stars, which contributes to up to $\sim 18$ per cent
of the initial total cluster mass after 15 Gyr), two-body relaxation,
interactions with the underlying, galactic tidal field, and dynamical
friction. The main results from these model runs, and of Vesperini's
(2000) comparison with a large sample of elliptical galaxies, are that
for a large number of host galaxy parameters, the mean mass and
dispersion do not differ significantly from their initial values,
although the fraction of surviving clusters, and therefore the cluster
disruption efficiency, does in fact vary significantly (see also
Vesperini 1998).

\subsection{Relevance to the M82 B cluster system}
\label{relevance.sec}

We will now consider the observational parameters of the M82 B cluster
system in the context of these evolutionary scenarios. We first need
to establish the relevance of these models for the interpretation of
the M82 B cluster system, however.

First, clusters on elliptical orbits are expected to disrupt more
quickly than those on circular orbits with the same apogalactic
distance (Baumgardt 1998; Baumgardt \& Makino 2003, Wilkinson et al.
2003). However, Vesperini's (1998, 2000) assumption that all clusters
are orbiting the galactic centre on circular orbits does not introduce
severe complications. M82 B is located between $\sim 500$ pc and 1 kpc
from its galactic centre, where the rotation curve of the H{\sc i} gas
has reached a constant velocity of $\sim 140$ km s$^{-1}$ (Wills et
al. 2000). Although at an age of $\sim 1$ Gyr the cluster system is
$\sim 25-50$ rotation periods old, the region is still spatially
closely confined. Thus, any deviation from circular velocities -- and
therefore any time dependence of the gravitational potential of the
host galaxy -- will have been experienced in a self-similar fashion by
the cluster system as a whole, and will not have propagated
differentially into the system.

Secondly, M82 is clearly not elliptical (nor a large spiral), but an
irregular galaxy. However, the main effect of constraining the shape
of the host galaxy to resemble an isothermal ``elliptical'' is that
the cluster system is evolving in a smooth gravitational potential.
Despite their appearance, the conditions governing the M82 B cluster
system (i.e., gravitational pull, cluster velocity distribution) are
remarkably similar to the simplified scenario of Vesperini's (1998,
2000) models.

Despite M82 being part of an interacting system, containing at least
six member galaxies (Yun 1999; including the three most massive
galaxies M81, M82 and NGC 3077), the time dependence of the
gravitational potential {\it caused by the gravitational interactions}
felt by M82 B is negligible. The mass of the system is dominated by
M81's gravitational potential, with a mass of $M_{\rm M81} \sim 20
\times 10^{10}$ M$_\odot$, while the mass of NGC 3077 is similar to
that of M82 ($M_{\rm NGC 3077} \sim 1 \times 10^{10}$ M$_\odot$,
$M_{\rm M82} \sim 2.5 \times 10^{10}$ M$_\odot$; Brouillet et
al. 1991, Yun 1999). However, even at perigalacticon, when M81
approached M82 to $\sim 25$ kpc some 220 to 510 Myr ago [Yun (1999),
Brouillet et al. (1991), respectively; i.e., well after the onset of
the burst of cluster formation], the absolute gravitational pull
exerted by the mass of M81 was negligible compared to the self-gravity
of M82. At the {\it onset} of the burst of cluster formation, when the
two galaxies were much farther apart, the effect was even smaller. We
conclude, therefore, that we can approximate the gravitational
potential felt by M82 B in a time-independent fashion, dominated by
the mass of M82 inside the radius of M82 B.

Furthermore, the very high mean density in the region, of $\langle
\rho \rangle \sim 2.5$ M$_\odot$ pc$^{-3}$, or $\log \langle \rho
\rangle ({\rm M}_\odot {\rm pc}^{-3}) \sim 0.4$ (de Grijs et
al. 2003a), has been instrumental in accelerating the cluster
disruption processes. Based on an assumed power-law IMF with $\alpha =
-2$ (de Grijs et al. 2003b) and a mass dependence of the disruption
time following $t_{\rm dis} \propto M_{\rm cl}^{0.62}$, as suggested
by observations and theory (see, e.g., Lamers et al. 2005a), we
derived a cluster disruption time-scale for M82 B of $t_{\rm dis} = 30
\times (M_{\rm cl} / 10^4 {\rm M}_\odot)^{0.62}$ Myr from the CMF (de
Grijs et al. 2003a). In this simple estimate, the disruption was
assumed to be instantaneous at an age $t_{\rm dis}$; this
simplification may have resulted in an underestimate of the disruption
time-scale (see Gieles et al. 2005). Therefore, the characteristic
disruption time-scale of the M82 B clusters might in fact be somewhat
longer. However, even with such a correction, this is the shortest
disruption time-scale known in any disc (region of a) galaxy (de Grijs
et al. 2003a).

\subsection{Constraints and implications from the M82 clusters}
\label{implications.sec}

Having established that Vesperini's (1998, 2000) models that start
from a log-normal CMF are indeed relevant in the context of the M82 B
intermediate-age cluster system, we will now discuss the implications
of the observational CLF and CMF parameters for this scenario.

After an initial rapid decrease in the mean cluster mass -- caused by
mass loss due to normal stellar evolution in the first $\sim 1$ Gyr
(Vesperini 2000; Baumgardt \& Makino 2003) -- the mean cluster mass is
expected to remain constant to within $\sim 0.025$ dex in $\langle
\log (M_{\rm cl} / {\rm M}_\odot) \rangle$, for an underlying galaxy
mass on the order of the mass of M82 (inside the radius of M82
B). This implies that the peak in the CMF (Fig. \ref{turnover.fig}b)
will likely remain constant for a Hubble time, within the
observational uncertainties [see Vesperini's (2000) Fig. 11; see also
Vesperini's (1998) discussion on the equilibrium CMF], irrespective of
the cluster disruption time-scale in this galaxy. The short disruption
time-scale in M82 B, if correct (see Section \ref{m82tdis.sec}), will
simply deplete the star cluster system at an accelerated rate compared
to galaxies with longer characteristic cluster disruption time-scales.

One of the main pieces of observational evidence in support of a
scenario in which the initial CMF in M82 B may have been log-normal is
therefore the fact that the turn-over in the M82 B CMF is observed for
an intermediate-age cluster system as young as $\sim 1$ Gyr, with
characteristic parameters (both the mean mass and the dispersion in
mass) very similar, if not identical, to the CMFs and CLFs in old GC
systems in the Milky Way, M31, M87 and old elliptical galaxies (e.g.,
Harris 2001). This would be very difficult to achieve if the initial
CMF had been closely approximated by a power-law (see Vesperini 1998,
2001). In the latter case, the mean mass will increase significantly
by the time the M82 B cluster system reaches an age similar to the
Galactic GCs, while the dispersion in mass decreases, so that one must
conclude that the M82 B CLF and CMF may not evolve into universal
distributions.

In addition, Vesperini (1998) showed that {\it any} reasonable
log-normal initial CMF will evolve towards the shape and parameters of
the equilibrium CMF, i.e., $\langle \log(M_{\rm cl}/{\rm
M}_\odot)\rangle \simeq 5.02$ (including the effects of disc shocking)
and $\sigma \simeq 0.67$. The speed at which this will occur depends
on the initial deviation of the system from the equilibrium
CMF. Therefore, the fact that for M82 B we observe $\langle
\log(M_{\rm cl}/{\rm M}_\odot)\rangle = 5.1 \pm 0.1 $ and $\sigma
\simeq 0.5$, at an age of $\sim 1$ Gyr, implies that the M82 B initial
CMF {\it must} have had a mean mass very close to that of the
equilibrium CMF [see Vesperini's (1998) Fig. 12]\footnote{We note that
under the assumption of a log-normal initial CMF, the NGC 1316 initial
CMF must have been characterised by a very low mean mass if it were to
evolve to its current mean mass of $\langle \log(M_{\rm cl}/{\rm
M}_\odot)\rangle \simeq 4.0$ at an age of $\sim 3$ Gyr, and it is
unlikely that this system will attain a Milky Way-type mean GC mass
over a Hubble time. In addition, none of the Vesperini models allow
for mean masses as low as those implied by the Goudfrooij et
al. (2004) results for NGC 1316.}.

Finally, Vesperini's (2000) Fig. 13 allows us to take the opposite
approach: if we are currently probing the {\it final} CMF, then the
{\it initial} CMF must have been characterized by a mean mass that was
only slightly larger than the present mean mass, and in fact still
within the observational uncertainties. This is a robust result, and
holds for gravitational potentials associated with host galaxies
spanning the entire observational range of masses and effective radii.

\section{A revised disruption time-scale for M82 B?}
\label{m82tdis.sec}

In de Grijs et al. (2003a), we used the distributions of the M82 B
cluster masses and ages to derive a cluster disruption time-scale for
this region of $t_{\rm dis} = 30 \times (M_{\rm cl} / 10^4 {\rm
M}_\odot)^{0.62}$ Myr, using the method developed by Boutloukos \&
Lamers (2003). Lamers et al. (2005a) used a similar approach to derive
the equivalent cluster disruption time-scales of the solar
neighbourhood, the Small Magellanic Cloud, and of selected regions in
M33 and M51. They compared their time-scales with those derived from a
number of $N$-body simulations in the parameter space defined by the
ambient density and the disruption time-scale. They found that
\begin{equation}
t_{\rm dis} = C_{\rm env} (M_{\rm cl}/10^4 {\rm M}_\odot)^{0.62}
(\rho_{\rm amb}/{\rm M}_\odot {\rm pc}^{-3})^{-0.5} \quad ,
\label{rhoamb.eq}
\end{equation}
where $\rho_{\rm amb}$ is the ambient density of the environment in
which the clusters move, and $C_{\rm env} \simeq 300$ to 800 Myr. This
predicted relation agrees with the empirically derived cluster
disruption time-scales in the SMC, M33, and the solar neighbourhood,
but not for M51 which has a significantly shorter disruption
time-scale.

We derived a rough estimate for the ambient density of M82 B in de
Grijs et al. (2003a), $\langle \rho \rangle \sim 2.5 {\rm M}_\odot
{\rm pc}^{-3}$, or $\log \langle \rho \rangle ({\rm M}_\odot {\rm
pc}^{-3}) \sim 0.4$, although with large uncertainties (see Section
\ref{relevance.sec}). If we nevertheless add these values for M82 B to
Fig. 4 in Lamers et al. (2005a), we see immediately that our new data
point for M82 B also lies well below the disruption lines predicted by
the two independently developed $N$-body simulations by Baumgardt \&
Makino (2003) on the one hand, and Portegies Zwart and colleagues on
the other (see Lamers et al. 2005b). This large discrepancy, of a
factor of $\sim 16$--17 in disruption time between the Baumgardt \&
Makino (2003) prediction and our result from de Grijs et al. (2003a),
prompted us to reconsider the assumptions on which we had based our
estimate. It is unlikely that we have underestimated the already
significant ambient density in M82 B by {\it several} orders of
magnitude. In fact, we believe this high ambient density estimate to
be an upper limit, as we will discuss below. Is it therefore possible
that we may have underestimated the disruption time-scale in M82 B?

Although we quote an uncertainty of about a factor of 2 in the
characteristic disruption time-scale, Fig. 7 of de Grijs et al.
(2003a) implies that our estimate of $t^4_{\rm dis}$ is a firm lower
limit, and seems to rule out a much shorter characteristic disruption
time-scale. 

However, our density estimate is a back-of-the-envelope guess with
uncertainties of about an order of magnitude. In fact, we used a
$V$-band mass-to-light (M/L) ratio of about 0.5--1 to obtain this
density estimate, but in de Grijs et al. (2001) we provided evidence
that the disk of M82 B shows active star formation until about 20--30
Myr ago. At an age of 25 Myr, the $V$-band M/L ratio $\sim 0.1$, i.e.,
a factor of 5--10 lower than what we used for our density
estimate. This reassessment, combined with the realisation that M82's
velocity curve (Wills et al. 2000) at the radius of region B indicates
a $\sim 3$ times smaller ambient density, $\rho_{\rm amb}$, confirms
our suspicion that the value used for $\rho_{\rm amb}$ is an upper
limit. Moreover, if we employ the 5--10 times lower $V$-band M/L ratio
instead, the M82B data point shifts to a location very close to that
of M51 in Fig. 4 of Lamers et al. (2005a).

We note that the most crucial assumption underpinning our disruption
time-scale estimate is that of the initial cluster mass distribution.
For both the M82 B time-scale, and in Lamers et al. (2005a), we used
an initial power-law CMF. This is a good assumption for young cluster
systems (de Grijs et al. 2003c). In de Grijs et al. (2003a) we showed
that in order to produce a log-normal present-day CMF in M82 B as
observed from an initial power-law distribution, an extremely short
disruption time-scale is required. However, we noted in Section
\ref{implications.sec} that the observed CMF in M82 B resembles
Vesperini's (1998) (quasi-)equilibrium CMF relatively closely. Let us
therefore consider the implications of this close coincidence in
shape, combined with the mass dependence of the disruption time-scale,
for the time-scale on which a typical $\sim 10^4$ M$_\odot$ cluster is
expected to disrupt.

Our 100 per cent completeness limit, shown in Fig. \ref{turnover.fig}
as the vertical dashed lines, occurs at a cluster mass of $\log(
M_{\rm cl}/ {\rm M}_\odot ) \simeq 4.4$ (Fig. \ref{turnover.fig}b).
The implication of our assumption of ``instantaneous disruption'' is
that $t_{\rm dis}^4 \lesssim 10^9$ yr, i.e., less than the present age
of the clusters formed simultaneously in the burst of cluster
formation. Therefore, we conclude that if the initial CMF in M82 B
were log-normal, we cannot constrain the characteristic disruption
time-scale to better than $t_{\rm dis}^4 \lesssim 10^9$ yr.

\section{Discussion}
\label{discussion.sec}

We randomly distributed 120,000 clusters, following two distinct mass
distributions: a power-law mass spectrum ${\rm d}N \propto M_{\rm
cl}^{\alpha} {\rm d}M_{\rm cl}$ and a log-normal mass function ${\rm
d}N/{\rm dlog}~M_{\rm cl}$. Lamers et al. (2005b) have shown that the
decreasing mass of a cluster can be described to a very high accuracy
as
\begin{equation}
\frac{M_{\rm cl}(t)}{M_{\rm i}}=\left\{ \bigl[ \mu_{{\rm
ev}}(t)\bigr]^\gamma - \frac{\gamma t}{t_{\rm dis}}
\right\}^{1/\gamma}\,,
\label{cl_mass_evol_L.eq}
\end{equation}
where $M_{\rm cl}(t)/M_{\rm i}$ is the mass of a cluster with initial
mass $M_{\rm i}$ that is still bound at an age $t$. In this equation,
$\mu_{\rm ev}(t)$ is the fractionary mass decrease of the cluster
because of stellar evolution only. The temporal evolution of $\mu_{\rm
ev}(t)$ is given by Eqs (2) and (3) of Lamers et al. (2005b), which
match the predictions of the GALEV SSP models very accurately. We
adopt $\gamma = 0.62$ (Boutloukos \& Lamers 2003; Baumgardt \& Makino
2003; Lamers et al. 2005b). The cluster disruption time-scale $t_{\rm
dis}$ depends on the initial mass $M_{\rm i}$ of the cluster and on
the ambient density $\rho_{\rm amb}$ as given by Eq.
(\ref{rhoamb.eq}). We will consider the disruption time-scale derived
by Baumgardt \& Makino (2003; their Eq. 10) based on a large set of
$N$-body simulations, taking into account the combined effects of
stellar evolution, two-body relaxation and the external tidal
field. In addition, we also investigate the disruption time-scale
obtained by de Grijs et al. (2003a), which is -- at the average
ambient density of M82 B, $\rho_{\rm amb} = 2.5$ M$_\odot$ pc$^{-3}$
(see Section \ref{m82tdis.sec}) -- $\sim 16 \times$ shorter than the
$N$-body simulation estimate. In order to take into account dynamical
friction, clusters with orbital decay time-scales (e.g., Binney \&
Tremaine 1994) shorter than a given time $t$ are removed from the
cluster system at that time. Owing to the intermediate age of the M82
B cluster system, the impact of dynamical friction proves negligible,
however. Finally, we normalised the mass distributions evolved to an
age of 1 Gyr to the observed number of clusters, i.e., the subsample
of the 42 clusters with the most accurately determined ages (see
Fig. \ref{turnover.fig}).

\begin{figure}
\psfig{figure=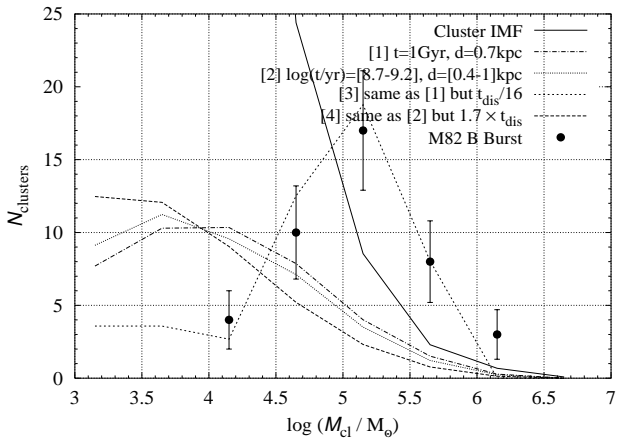,width=8.6cm}
\caption{\label{powerlaw.fig}Evolution of an initial power-law CMF
(solid line) for 1 Gyr, assuming both the short disruption time-scale
of $t_{\rm dis}^4 \sim 30$ Myr ([3], short dashed line; de Grijs et
al. 2003a,b) and the $\sim 16 \times$ longer time-scale based on
Baumgardt \& Makino's (2003) $N$-body simulations ([1,2] dotted and
dot-dashed lines).  The dotted and dot-dashed lines show the small
differences between a coeval (exactly) 1 Gyr-old cluster population in
M82 B, located at a galactocentric distance of (exactly) 0.7 kpc
($\rho_{\rm amb} = 2.5$ M$_{\odot}$ pc$^{-3}$ and $t_{\rm dis}^4 \sim
0.5$ Gyr), versus the scenario in which the clusters show a uniform
age spread over $8.7 \le \log( {\rm Age/yr} ) \le 9.2$ and are
distributed uniformly across the region ($0.4 \le (D/{\rm kpc}) \le
1.0$). The dashed line ([4]) illustrates the slight shift of the
evolved CMF turnover towards lower mass in case of a more realistic
ambient density of 0.8 M$_\odot$ pc$^{-3}$, corresponding to $t_{\rm
dis}^4 \sim 0.8$ Gyr. The observational data points and their
Poissonian error bars (see Fig. \ref{turnover.fig}) are also
included.}
\end{figure}

\begin{figure}
\psfig{figure=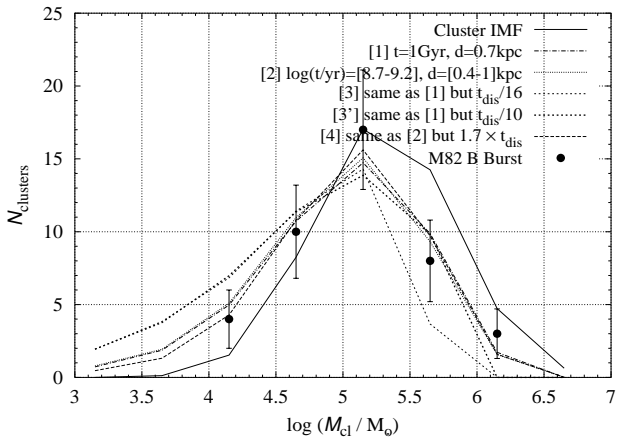,width=8.6cm}
\caption{\label{lognormal.fig}As Fig. \ref{powerlaw.fig}, but for an
initial log-normal CMF.}
\end{figure}

To assess the robustness of the evolved CMFs with respect to the age
and spatial distributions of the clusters, we considered the following
cases. First (cases [1,3] in Figs. \ref{powerlaw.fig} and
\ref{lognormal.fig}), all clusters were assumed to be 1 Gyr old and
located at the same galactocentric distance of $D = 0.7$ kpc.
Consequently, they are all characterized by the same ambient density
($\rho_{\rm amb} \simeq 2.5$ M$_\odot$ pc$^{-3}$), and thus by the
same disruption time-scale. Secondly (case [2] in Figs.
\ref{powerlaw.fig} and \ref{lognormal.fig}), we considered the case of
a cluster system characterized by uniform distributions in age and
galactocentric distance. Following our definition of the burst of
cluster formation (see Fig. \ref{turnover.fig}), the lower and upper
limits of the age distribution are $\log(t/{\rm yr}) = 8.7$ and 9.2,
respectively. As for the spatial distribution, clusters were assumed
to be distributed uniformly in galactocentric distance across the
region ($0.4 \le (D/{\rm kpc}) \le 1.0$), the radial extent of M82
B. In this case, the cluster system probes a range of ambient
densities and, therefore, of characteristic disruption time-scales. We
assume that the radial profile of the ambient density is that of a
singular isothermal sphere, $\rho_{\rm amb} \propto D^{-2}$, and that
$\rho_{\rm amb}(D=0.7\,{\rm kpc})=2.5$ M$_\odot$ pc$^{-3}$.

The initial power-law CMF is characterised by a slope of $-2$ (see de
Grijs et al. 2003c for a review). Fig. \ref{powerlaw.fig} shows the
corresponding evolved mass distributions, using both the disruption
time-scale determined by de Grijs et al. (2003a; case [3]), and the
$\sim 16 \times$ longer time-scale suggested by Baumgardt \& Makino's
(2003) $N$-body simulations (case [1]). We also show the small
differences between a coeval (exactly) 1 Gyr-old cluster population,
located at a galactocentric distance of (exactly) 0.7 kpc (case [1]),
and the scenario in which the clusters show a uniform age spread over
$8.7 \le \log( {\rm Age/yr} ) \le 9.2$ and are distributed uniformly
across the region ($0.4 \le (D/{\rm kpc}) \le 1.0$). This uniform
distribution in galactocentric distance corresponds to a number
density profile scaling as $D^{-2}$. We checked that the differences
caused by assuming radial density profiles following a fairly
arbitrary (and shallow) profile $\propto D^{-0.5}$, as well as a much
steeper number density distribution such as that of GCs in the
Galactic halo, $\propto D^{-3.5}$ (e.g., Zinn 1985), result in
identical evolved mass distributions, within the uncertainties. The
assumed spatial distribution of the clusters affects our results
negligibly, which is a natural consequence of the small radial extent
of M82 B. In view of the region's disturbed appearance and unique star
(cluster) formation history, neither a uniform nor a strongly radially
dependent initial density distribution can be ruled out {\it a priori}.

As expected, if we employ $t_{\rm dis}^4 = 30$ Myr (case [3]), the
resulting mass distribution is log-normal, and matches the observed
mass distribution (Fig. \ref{powerlaw.fig}) very closely. This is not
surprising, since the determination of this short disruption
time-scale was based on the assumption of an initial power-law CMF
with a slope of $-2$. [This shows that the assumption of instantaneous
disruption, adopted in de Grijs et al. (2003a,b) hardly affects the
determination of the disruption time-scales.]

If, instead, we use an initial power-law CMF as before, but now assume
that the longer disruption time-scale predicted by Baumgardt \&
Makino's (2003) $N$-body simulations is correct (cases [1,2]), the
evolved cluster mass distribution shows a (broader) peak, shifted to
lower masses by more than one order of magnitude. Moreover, we argued
in the previous section that our estimate of the ambient density is
likely to have been overestimated by at least a factor of 3. Since, in
an undisturbed tidal field of a galaxy with a logarithmic potential,
the disruption time-scale depends on the ambient density as $t_{\rm
dis} \propto \rho_{\rm amb}^{-0.5}$, we have also evolved a power-law
with a $1.7\times$ larger disruption time-scale (case [4] in
Fig. \ref{powerlaw.fig}); this factor of 1.7 allows for the
uncertainty in $\rho_{\rm amb}$ between 0.8 and 2.5 M$_\odot$
pc$^{-3}$. As expected, this longer, probably more realistic
disruption time-scale gives rise to a turnover located at a cluster
mass smaller than that derived in cases [1] and [2], thus
strengthening the discrepancy between the evolved model and observed
CMFs. Therefore, {\it if} we assume that Baumgardt \& Makino's (2003)
$N$-body simulations predict approximately the appropriate cluster
disruption time-scale for M82 B, Fig. \ref{powerlaw.fig} shows that
the observed cluster mass distribution {\it cannot} be retrieved from
an initial power-law CMF.

We will now approach this issue starting from an initial log-normal
CMF. We have assumed that the initial log-normal CMF matches that of
the almost universal mass distribution of old GC systems in the local
Universe (and thus that of the theoretical (quasi-)equilibrium CMF of
Vesperini 1998). If the Baumgardt \& Makino (2003) results apply
(i.e., $t_{\rm dis}^4 \sim 0.5{\rm -}0.8$ Gyr at an ambient density
typical of M82 B (i.e., $\rho_{\rm amb} \simeq 0.8{\rm -}2.5$
M$_\odot$ pc$^{-3}$), most of the clusters in the log-normal initial
CMF of Fig. \ref{lognormal.fig} will not yet have been significantly
affected by disruption. Thus, when we evolve this initial CMF to an
age of 1 Gyr, the CMF approximately retains its initial shape, as
shown in Fig. \ref{lognormal.fig}. This result holds irrespective of
the underlying cluster age and distance distributions and irrespective
of the average ambient density assumed (i.e., the evolved CMFs derived
in cases [1,2,4] are identical well within the observational
uncertainties). The main difference between the initial and evolved
CMFs is therefore caused by the effects of stellar evolution: the
shift of the peak of the distribution by $\Delta \log(M_{\rm cl}/{\rm
M}_\odot) \sim -0.15$ is the result of up to 25 per cent of stellar
evolutionary mass loss. With the short disruption time-scale of de
Grijs et al. (2003a), the final distribution is somewhat more depleted
in high-mass clusters (case [3] in Fig. \ref{lognormal.fig}). We also
show the evolved 1 Gyr-old CMF assuming a slightly longer disruption
time-scale $t_{\rm dis}^4 \sim 50$ Myr (case [3$'$]). At an ambient
density of 2.5 M$_\odot$ pc$^{-3}$, this is $10\times$ shorter than
that derived from Baumgardt \& Makino 's (2003) $N$-body simulations.
The predicted CMF still matches the observed distribution
satisfactorily. Therefore, we note that when starting from an initial
log-normal CMF, the evolved mass distributions match the observed
distribution in M82 B (Fig. \ref{turnover.fig}b) fairly closely, and
that this result holds {\it for a wide range of disruption
time-scales}.

To assess the robustness of the results presented in Figs.
\ref{powerlaw.fig} and \ref{lognormal.fig}, we have also evolved the
log-normal and power-law initial CMFs using Eq. (12) of Baumgardt \&
Makino (2003). A detailed comparison shows that the results derived
based on Eq. (\ref{cl_mass_evol_L.eq}) above appear robust: the shape
of the Gaussian CMF is unaffected by the 1 Gyr-long evolution in
either case, and the turnover of the evolved power-law CMF is
discrepant with the observed peak by more than one order of magnitude.

\subsection{Unphysically high initial densities?}

We note, however, that Baumgardt \& Makino's (2003) $N$-body
simulations were performed assuming a smooth underlying tidal field;
they do not include the effects of external perturbations such as
those caused by encounters with giant molecular clouds. As a result,
their cluster disruption time-scale is therefore an upper limit. In
view of the uncertainties inherent to the precise disruption
time-scale governing M82 B we cannot use this analysis by itself to
distinguish conclusively between the log-normal vs. power-law initial
CMF. The key question is then whether the combination of (i) an
initial power-law CMF, (ii) the present number (and mass) of 1 Gyr-old
clusters, and (iii) the very short disruption time-scale of $\sim 30$
Myr for a $10^4$ M$_\odot$ cluster can be accommodated in a physically
realistic scenario.

Starting from a power-law initial CMF with masses between $10^3$ and
$3 \times 10^6$ M$_\odot$, the ratio of the final (i.e., at an age of
1 Gyr) to initial number of clusters is $F_N \simeq 5 \times 10^{-4}$
if $t_{\rm dis}^4 \sim 30$ Myr. In this case, the 1 Gyr-old clusters
considered here are the survivors of an initial population of $\simeq
8 \times 10^{4}$ clusters. However, this is a lower limit since these
42 clusters constitute a subsample of the M82 B cluster population,
i.e., those located at the ``surface'' of the region {\it and} for
which reliable age estimates could be obtained. For an initial cluster
mass range with a lower limit of $10^4$ M$_\odot$, the initial number
of clusters is significantly lower, i.e., some 8,000, but still very
large for the spatially confined M82 B region. The ratio $F_{\rm M}$
of the final to the initial mass in clusters is $\lesssim 1$ per
cent. The present mass of the observed cluster system is {\it at
least} $\sim 10^7$ M$_\odot$ (and likely much more considering that we
have only sampled the outer surface of the region). This implies,
therefore, that the initial mass in (bound, long-lived) clusters alone
must have been on the order of $10^9$ M$_\odot$, confined to a
three-dimensional volume of $\lesssim 5 \times 10^7$ pc$^3$ (de Grijs
et al. 2003a,b). The initial mean density {\it in bound clusters
alone} must therefore have been $\gtrsim 20$ M$_\odot$ pc$^{-3}$, if
the initial CMF were a power-law distribution. This is at least an
order of magnitude higher than the current {\it total stellar} density
in M82 B, as well as in the actively cluster-forming centre of M51
(Lamers et al. 2005a). Since the mass in clusters generally only
comprises a few per cent of the total mass in disc galaxies, up to
about 30 per cent in dense starburst regions like M82 B\footnote{If a
cluster is in equilibrium with its environment, as we argued for M82 B
in de Grijs et al. (2003a), one can estimate that $\rho_{\rm
cl}/\rho_{\rm amb} \simeq 3$ for a standard Roche solution by assuming
that the clusters are characterized by a King (1966) profile, and that
its tidal radius equals the Jacobi radius of the host galaxy's tidal
field (see Binney \& Tremaine 1994; Lamers et al. 2005a).}, it follows
that the initial total stellar density required may be as high as
$\sim 60$ M$_\odot$ pc$^{-3}$. Such densities are physically
unrealistic in disc regions of ``normal'' galaxies, even in dense
starburst regions. We note in passing that these calculations refer to
the (initially) bound clusters only; if unbound clusters were
included, the expected initial mean density would be even higer.

Therefore, we conclude that our observations of the present M82 B CMF
are inconsistent with a scenario in which the 1 Gyr-old cluster
population originated from an initial power-law mass distribution.
Note that this applies both to the very short disruption time-scale of
$\sim 30$ Myr, as well as to the longer time-scale based on the
Baumgardt \& Makino (2003) results, for which we concluded above that
the resulting present-day CMF would peak at much lower masses than
observed.

\begin{figure}
\psfig{figure=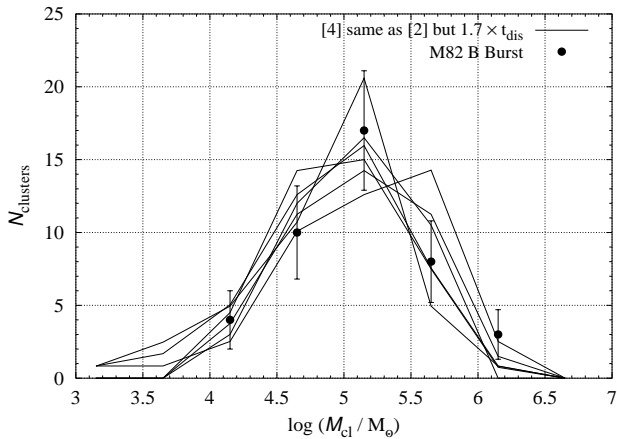,width=8.6cm}
\caption{\label{lognormal_N60.fig} As case [4] in
Fig.~\ref{lognormal.fig}, but starting from 60 clusters. Evolved CMFs
are presented for various random realisations.}
\end{figure}

For a log-normal initial CMF combined with the Baumgardt \& Makino
(2003) disruption time-scale ($t_{\rm dis}^4 \simeq 0.5{\rm -}0.8$
Gyr), most of the clusters survive the 1 Gyr-long evolution, i.e.
$F_N \simeq 0.9$. The initial and final numbers of clusters are thus
very similar. In order to explore whether the good match between the
evolved Gaussian model CMF and the observed distribution actually
depends on the small initial number of clusters implied by this
survival rate (i.e., whether the mass distribution of the actual data
points may be affected significantly by small-number statistics rather
than physical effects), we have run simulations starting from 60
clusters (i.e., the approximate number of clusters expected to have
been formed initially in this scenario), randomly drawn from the
Gaussian initial CMF. The results are shown in Fig.
\ref{lognormal_N60.fig} for case [4], where we assume the disruption
time-scale predicted by $N$-body simulations ($t_{\rm dis}^4 \simeq
0.8$ Gyr). For various random seeds, the predicted cluster mass
distributions match the observed CMF satisfactorily, showing that the
effects owing to small-number statistics are minimal, and unimportant
with respect to our overall conclusions.

It appears, therefore, that on the grounds of both our observational
data and the theoretical arguments presented in the previous sections,
the initial mass distribution of the M82 B clusters surviving past the
``infant mortality'' epoch (i.e., the first few Myr in which unbound
low-mass clusters are dispersed; e.g., Boily \& Kroupa 2003; Vesperini
\& Zepf 2003; Whitmore 2004; Bastian et al. 2005; Mengel et al. 2005;
see also Tremonti et al. 2001) must have been closely matched by a
log-normal distribution.

In fact, the presence of a large excess (up to 70--90 per cent;
Whitmore 2004; Mengel et al. 2005) of presumably unbound clusters at
ages below $\sim 10$ Myr in the Antennae system (Whitmore 2004, his
fig. 2; Mengel et al. 2005, their fig. 10) and M51 (Bastian et
al. 2005, their fig. 10) could, in principle, provide further limits
on the initial CMF of both the bound and unbound clusters. A seven to
nine-fold increase of unbound clusters at very early times, as implied
by these observational studies, although of low mass in general, would
boost initial stellar density levels to truly unphysical numbers if
the bound, longer-lived clusters were formed following a power-law
mass-number scaling. For the initial log-normal CMF scenario, the
resulting initial densities could be used to place limits on the total
mass (and possibly the number, if the mass distribution were known) in
unbound star clusters. However, at this point the observational data
are statistically insufficiently robust in terms of excess cluster
numbers (similar data are needed for larger numbers of cluster
populations), while the masses of these young unbound cluster
populations are as yet poorly determined, so that any (statistical)
extrapolations to other galaxies are as yet unwarranted.

\section{Summary and Conclusions}
\label{conclusions.sect}

In this paper, we start from the robust detection of de Grijs et
al. (2003a,b) of an approximately log-normal CMF for the 1 Gyr-old,
intermediate-age star cluster system in M82 B, and explore whether we
can constrain the shape of the {\it initial} distribution of cluster
masses.

In particular, we investigate whether the most likely initial CMF was
more similar to either a log-normal or a power-law distribution, by
taking into account the dominant evolutionary processes (including
stellar evolution, and internal and external gravitational effects)
affecting the mass distributions of star cluster systems over
time-scales of up to $\sim 1$ Gyr in the presence of a realistic
underlying gravitational potential. The M82 B cluster population
represents an ideal sample to test these evolutionary scenarios for,
since it is a roughly coeval intermediate-age population in a
spatially confined region. For such a coeval population, the
observational selection effects are very well understood (de Grijs et
al. 2003a,b), while the dynamical cluster disruption effects are very
similar for this entire population.

After considering the gravitational effects and geometry of M82
itself, its starburst region B, and its position in the M81/M82/NGC
3077 group of interacting galaxies, we conclude that we can
approximate the gravitational potential felt by M82 B in a
time-independent fashion, dominated by the mass of M82 inside the
radius of M82 B. In such a static gravitational potential, Vesperini
(1998) shows conclusively that there exists a particular CMF of which
the initial mean mass, width and radial dependence remain unaltered
during the entire evolution over a Hubble time. In fact, the mean mass
and width of {\it any} initial log-normal CMF tends to evolve towards
the values for this equilibrium CMF.

Thus, the fact that for M82 B we observe $\langle \log( M_{\rm cl} /
{\rm M}_\odot )\rangle = 5.1 \pm 0.1 $ and $\sigma \simeq 0.5$, at an
age of $\sim 1$ Gyr, implies that the M82 B initial CMF {\it must}
have had a mean mass very close to that of the equilibrium CMF. If the
presently observed M82 B CMF is to remain unchanged for a Hubble time,
so that we are currently probing the {\it final} CMF, then the {\it
initial} CMF must have been characterized by a mean mass that was only
slightly larger than the present mean mass. This is a robust result,
and holds for gravitational potentials associated with host galaxies
spanning the entire observational range of masses and effective radii.

From our detailed analysis of the expected evolution of CMFs starting
from initial log-normal and initial power-law distributions, we
conclude that our observations of the M82 B CMF are inconsistent with
a scenario in which the 1 Gyr-old cluster population originated from
an initial power-law mass distribution. This applies to a range of
characteristic disruption time-scales, from $t_{\rm dis}^4 \sim 30$
Myr to the $\sim 16-30 \times$ longer time-scale resulting from
Baumgardt \& Makino's (2003) $N$-body simulations. Our conclusion is
supported by arguments related to the initial density in M82 B, which
would be unphysically high if the present cluster population were the
remains of an initial power-law distribution (particularly in view of
the effects of cluster ``infant mortality'', which require large
excesses of low-mass unbound clusters to be present at the earliest
times).

De Grijs et al. (2003c) showed that the CMFs of YSCs in many different
environments are well approximated by power laws with slopes $\alpha
\simeq -2$. However, except for the intermediate-age cluster systems
in M82 B (de Grijs et al. 2003a,b) and NGC 1316 (Goudfrooij et
al. 2004), the {\it expected} turn-over mass (based on comparisons
with present-day GC systems and taking evolutionary fading into
account) in most YSC systems observed to date occurs close to or below
the observational detection limit, simply because of their greater
distances and shallower observations. As such, the results presented
here and those summarised in de Grijs et al. (2003c) are not
necessarily at odds with each other, but merely hindered by
observational selection effects.

\section*{acknowledgments} RdG acknowledges support from The
British Council under the {\sl UK--Netherlands Partnership Programme
in Science}, and from the Nuffield Foundation through a New Lecturers
in Science, Engineering and Technology Award, and hospitality at the
Astronomical Institute of Utrecht University on several visits. GP
acknowledges research support from a Marie Curie Intra-European
Fellowship within the sixth European Community Framework Programme. We
are grateful for research support and hospitality at the International
Space Science Institute in Bern (Switzerland), as part of an
International Team programme. This research has made use of NASA's
Astrophysics Data System Abstract Service.

\end{document}